\pdfoutput=1

\documentclass{iaus}
\usepackage{graphicx}
\usepackage{url}

\title[High-energy emission from galaxies] 
{High-energy emission from galaxies: \\ 
the star-formation/gamma-ray connection}

\newcommand{\g}{$\gamma$}
\newcommand{\hess}{H.E.S.S.}

\author[Stefan~Ohm \& Jim~Hinton] 
{Stefan~Ohm$^{1,2}$\thanks{SO acknowledges the support of the
    Humboldt foundation by a Feodor-Lynen research fellowship.}  \and
  Jim~Hinton$^1$} \affiliation{ $^1$X-ray and Observational Astronomy
  Group,
  Department of Physics and Astronomy, \\
  University of Leicester, LE1~7RH, UK\\
  email: {\tt jim.hinton@le.ac.uk} \\[\affilskip]
  $^2$School of Physics and Astronomy,
  University of Leeds, LS2~9JP, UK\\
  email: {\tt stefan.ohm@le.ac.uk} }

\pubyear{2011} 
\volume{284}  
\pagerange{1--12}
\jname{The Spectral Energy Distribution of Galaxies}
\editors{R.J. Tuffs \&  C.C.Popescu, eds.}
\begin{document}

\maketitle

\begin{abstract}
  The impact of non-thermal processes on the spectral energy
  distributions of galaxies can be dramatic, but such processes are
  often neglected in considerations of their structure and
  evolution. Particle acceleration associated with high mass star
  formation and AGN activity not only leads to very broad band
  (radio-\g-ray) emission, but may also produce very significant
  feedback effects on galaxies and their environment. The recent
  detections of starburst galaxies at GeV and TeV energies suggest
  that \g-ray instruments have now reached the critical level of
  sensitivity to probe the connection between particle acceleration
  and star-formation in galaxies. In this paper we will try to
  summarise this recent progress, put it into a multi-wavelength
  context and also discuss the prospects for more precise and
  sensitive \g-ray measurements with the upcoming CTA observatory.
\end{abstract}

\firstsection 
\section{Introduction}

The most dramatic examples of high-energy emission from galaxies are
found in Active Galactic Nuclei (AGN), where particle acceleration
occurs in jets powered by supermassive black holes. In such systems
non-thermal emission is often seen across the whole electromagnetic
spectrum (in systems where the jets are aligned with the line-of-sight
to the observer -- commonly known as blazars). In the more common case
that jets are not aligned to the observer, non-thermal dominance
occurs only in the radio, X-ray and \g-ray domains. The high-energy
emission of these objects has been discussed by many authors (see
e.g. \cite{Wagner08} for a compilation of VHE \g-ray blazars), here we
focus on the less dramatic, but potentially very importent non-thermal
emission of ``inactive'' galaxies, where the spectral energy
distribution (SED) is dominated by stellar processes.

In the Milky Way (MW), we know that the population of
ultra-relativistic particles, known as cosmic rays (CRs), plays an
important dynamical role; being close to pressure equilibrium with
thermal gas and magnetic fields in the Interstellar Medium (ISM). Our
knowledge of this population has in the past come largely from direct
(local) measurements and from radio-synchrotron emission from the
energetically less important relativistic electrons, rather than the
nuclei that make up 99\% of the CRs.

The process of star formation (SF), particularly of massive stars, is
now known to lead to astrophysical particle acceleration, and hence
\g-ray emission, via a number of different objects and phenomena. The
well-established accelerators are associated with the end-products of
the massive-stellar lifecycle: supernova remnants (SNR) and pulsars
and their associated pulsar wind nebulae (PWNe). Over the last two
decades, measurements in the X-ray and \g-ray domains have shown that
SNRs are efficient particle accelerators (see e.g. \cite{Vink2011}),
with increasing evidence for the acceleration of nuclei as well as
electrons (e.g. \cite{HESS:W28,Fermi:W28}). Less secure are the
apparent association of \g-ray emission with colliding stellar winds
in binary systems (\cite{Farnier2011}) and the collective effect of
winds in clusters (see e.g. \cite{HESS:Wd1_11}).

Independent of the dominant contributor to the CR population, the
expectation from our own Galaxy is that all massive star-forming
regions should be associated with particle acceleration and hence
\g-ray emission, at some level. Indeed, recent progress in the field
of high-energy (HE) and very-high-energy (VHE) \g-ray astronomy,
driven by the satellite-borne instrument \emph{Fermi}-LAT and
ground-based VHE telescopes such as \hess\ and VERITAS, has led to the
identification of a growing population of high-energy-emitting
star-forming galaxies (including the nearby starburst galaxies M\,82
and NGC\,253).  We focus on these new developments below, discussing
in particular the case of starburst galaxies, and finally discuss the
prospects for the next-generation \g-ray mission CTA.

\section{Particle energy losses and \g-ray production}
\label{sec:losses}

Ultra-relativistic electrons and nuclei suffer energy losses by a
number of different mechanisms, which lead to the production of
photons from the radio to VHE \g-ray regime. Here we consider the
primary energy-loss mechanisms in turn, giving example loss-timescales
appropriate for the environment in starburst regions.  CR nuclei
predominantly lose energy in strong nuclear interactions with ambient
matter, producing $\pi^0$-decay \g-ray emission primarily above a few
hundred MeV. For a typical starburst region the (fairly
energy-independent) interaction timescale is $t_{pp}\approx
10^{5}\,(n/250\,{\rm cm}^{-3})^{-1}$ years, where $n$ is the average
ambient density in hydrogen atoms per cm$^3$, with an average energy
loss of $\sim$50\% per collision. For electrons a number of different
energy-loss mechanisms play a significant role. For the highest energy
electrons synchrotron and inverse Compton (IC) emission dominate with
energy loss timescales ($E/dE/dt$) of $t_{\mathrm{sync}}\approx
200\,(E/{\rm TeV})^{-1}$ years and $t_{\mathrm{IC}}\approx300\,(E/{\rm
  TeV})^{-1}$ years, for a starburst-like environment with an average
magnetic field strength of $250\mu$G, and a radiation field energy
density of $1000$\,eV\,cm$^{-3}$. At electron energies below a few GeV
Bremsstrahlung and (at the lowest energies) Coulomb losses become
increasingly important, impacting on the shape of the equilibrium
(i.e. cooled) electron spectrum and hence the spectrum of the
non-thermal emission (see Figure \ref{fig1} and Section 4 for more
details). In such an environment a 1\,TeV electron will produce IC and
synchrotron photons with typical energies of $\approx150$\,GeV (for
black-body target photons with $T=50$\,K) and $\approx20$\,keV,
respectively.

These timescales are so short that it is often very difficult for high
energy electrons to escape from star-forming regions. Nuclei, in
contrast may be removed by diffusion or bulk motion before significant
energy losses occur. Starburst galaxies often drive winds which carry
away particles from the nuclear region into the inter-galactic
medium. This advection of particles is determined by the wind speed
and its scale height, and can be characterised by an average residence
time of the particles in the nucleus -- which is around $2\times10^5$
years for the $\sim500$\,km\,s$^{-1}$ starburst wind of NGC\,253. More
details on energy loss and \g-ray production mechanisms in general can
be found in \cite{FelixBook,Longair11,Hinton09}. For more details on
the environment and high-energy emission of starbursts see for example
\cite{NGC253:Pag96,NGC253:Lacki11} and references therein.

\section{Non-thermal emission from Starburst galaxies}


The presence of high-energy electrons in the archetypal starburst
galaxies NGC\,253 and M\,82 was identified through low-frequency
radio-synchrotron emission measurements in the late 1970's
(e.g. \cite{Shimmins1973} and early 1980's (e.g. \cite{Laing1980}),
respectively. VHE \g-ray emission from these two objects has only
recently been reported by \cite{HESS:NGC253} and \cite{VERITAS:M82},
respectively. Both galaxies have also been detected in the HE domain
using data obtained in observations with the LAT instrument onboard
the \emph{Fermi} satellite (\cite{Fermi:NGC253M82}). Both starbursts
are right at the sensitivity limit of \emph{Fermi}-LAT and
ground-based VHE \g-ray instruments -- their detection required very
long exposure times and they comprise the weakest \g-ray source class
detected so far. Within large errors, the HE and VHE \g-ray spectrum
of M\,82 can be described by a single power-law in energy with photon
index $\Gamma\approx 2.3$. Similarly, the extrapolated HE \g-ray
spectrum of NGC\,253 ($\Gamma\approx 2.3$) is consistent with the
integrated VHE \g-ray flux reported by H.E.S.S. In the following, we
will outline the importance of different energy-loss processes of
nuclei and electrons in a typical starburst environment as discussed
in Section~\ref{sec:losses} and illustrate the effect on the SED of
changing key parameters of the system with a very simple model.

\begin{figure}
  \begin{center}
    \includegraphics[width=\textwidth]{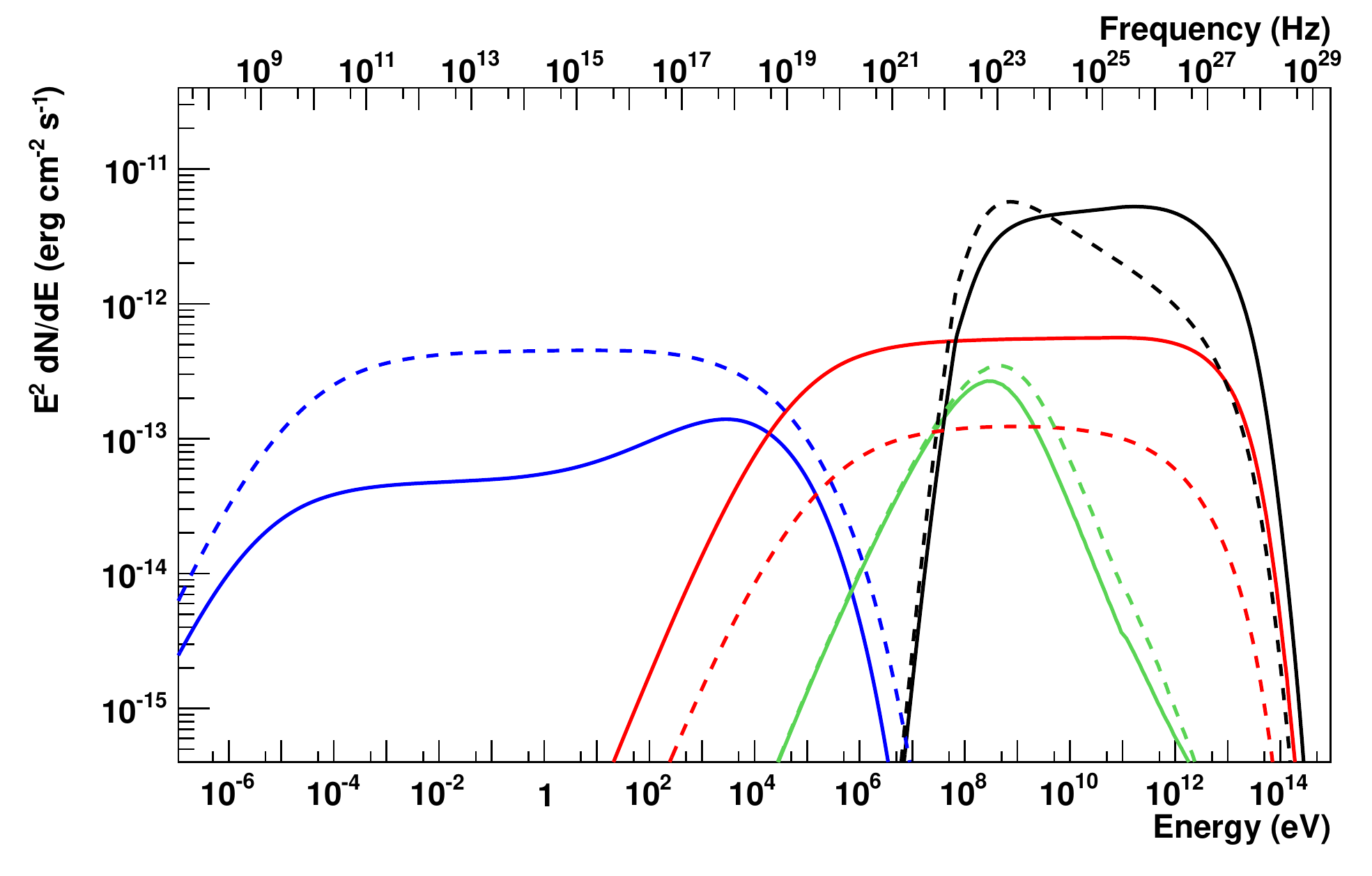} 
    \caption{Simple model SED for a representative starburst region at
      a distance of 3.5\,Mpc. Colours indicate the contributions of IC
      (red), synchrotron (blue), Bremsstrahlung (green) and
      $\pi^0$-decay (black) in a single-zone, time-dependent model for
      the continuous injection of electrons and protons over
      $2\times10^5$\,yr. For the solid lines (model 1), a magnetic
      field strength of $B=100\,\mu$G, a radiation field energy
      density of $U_{\mathrm{rad}}=2500$\,eV of a black-body with
      temperature 50\,K, and an average density n$_{\mathrm{H}}$ of
      250 particles per cm$^3$ are assumed. The energy input is
      $3\times10^{40}$\,erg\,s$^{-1}$ for electrons and an order of
      magnitude higher for protons. An injection spectrum index of
      $\alpha=2.0$ and maximum accelerated particle energy of
      $E_{\mathrm{max}} = 100$\,TeV are used for this model. Dashed
      lines (model 2) illustrate the effect of reducing $U_{\rm rad}$
      by a factor of 10 and doubling the magnetic field strength. A
      proton spectral index of 2.3 is used in this second case.}
    \label{fig1}
  \end{center}
\end{figure}

Fig.~\ref{fig1} shows a illustrative model SED for the particle
injection and cooling in an environment similar to that found in the
starburst nuclei of NGC\,253 and M\,82. Electrons and protons are
injected over the assumed particle residence time of $2\times10^5$
years. As this timescale is longer than the cooling timescales for
electrons in the system, this SED is representative of the system at
equilibrium (i.e. the present day SED -- given the typical starburst
lifetime of $\sim10^{7}$ years), provided that the escape probability
is energy-independent.  The IC, synchrotron and bremsstrahlung
emission for electrons is shown for ``typical'' and ``extreme'' values
of the radiation field energy density and magnetic field strength.
For model~1 a comparatively low $B$-field strength of 100\,$\mu$G and
rather high radiation field density of 2500\,eV\,cm$^{-3}$ are
assumed.  In this model a high-energy upturn in the synchrotron
spectrum can be seen -- a consequence of IC-dominated cooling in the
Klein-Nishina regime. In model~2 the magnetic field strength is
assumed to be $B=200$\,$\mu$G and $U_{\rm rad}=250$\,eV\,cm$^{-3}$,
and synchrotron losses always dominate over IC.  The energy input of
$3\times10^{41} (3\times10^{40})$\,erg\,s$^{-1}$ for protons
(electrons) represents an electron-to-proton ratio of 1/10 and has
been derived assuming one supernova explosion every decade and a
particle acceleration efficiency of 10\%. As the electron-to-proton
ratio at injection in our own galaxy is likely $\sim$1/100, this SED
illustrates that $\pi^{0}$-decay \g-rays are very likely to dominate
the SED at GeV energies. For electrons with energies much smaller than
$\approx 1$\,GeV, Coulomb cooling starts to become significant and
results in a hardening of the radio spectrum in the 100\,MHz to 1\,GHz
frequency range.

These qualitative results should be compared to the more sophisticated
modelling of non-thermal emission of starbursts in (for example)
\cite{NGC253:Pag96,NGC253:Domingo05,M82:Persic08} or
\cite{NGC253:Lacki11}.

\section{A comparison of NGC 253 and the Milky Way}

Now that \g-ray emission from external star-forming galaxies has been
established, we have the opportunity for the first time to compare the
high-energy emission of the MW to other systems and test our ideas on
relativistic particle production and transport.  As a very nearby
starburst with well-established radio, GeV and TeV emission, NGC\,253
is well suited to such a comparison.

\begin{figure}
\begin{center}
 \includegraphics[width=\textwidth]{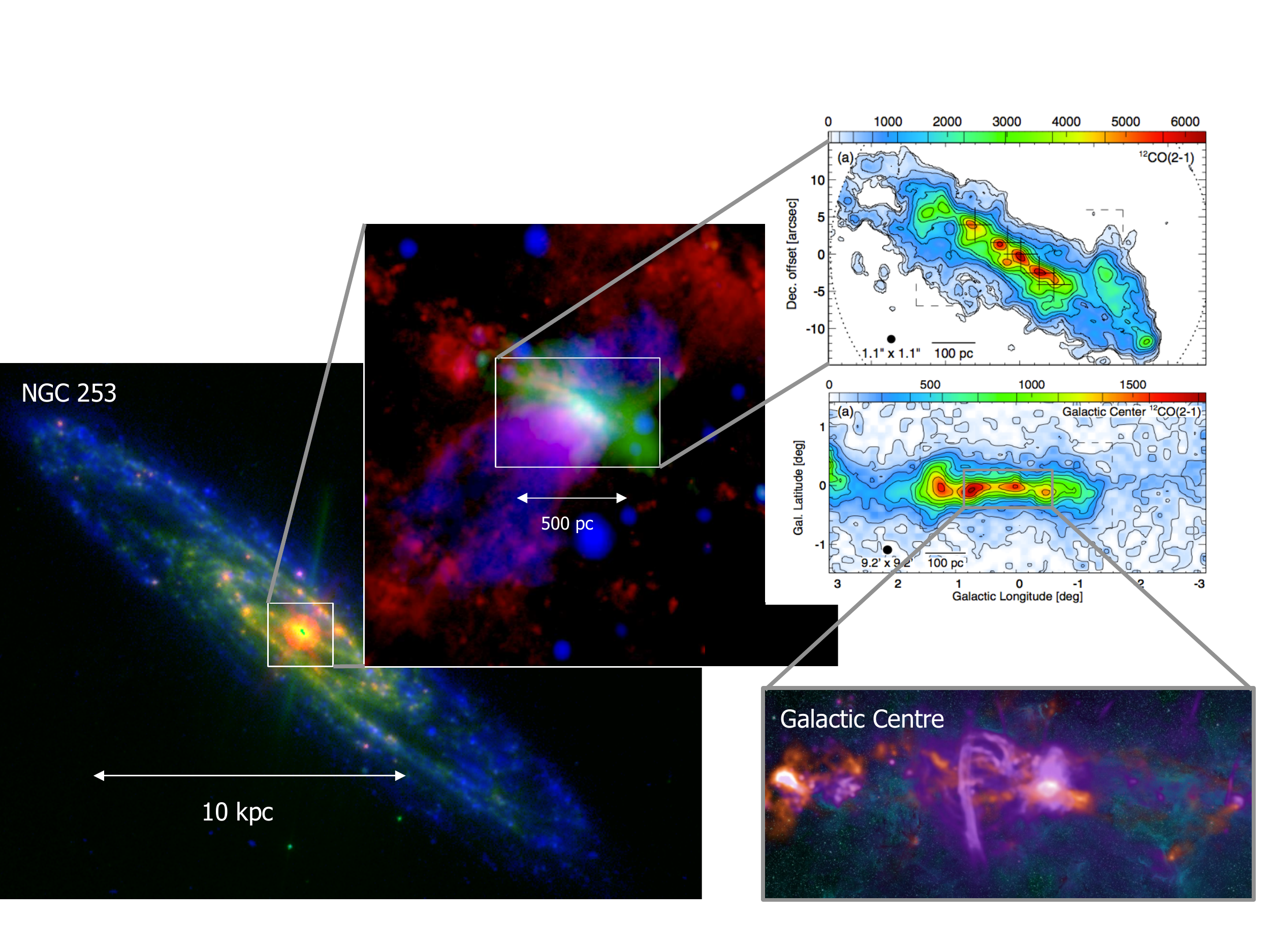} 
 \caption{Composite Spitzer and GALEX image of NGC\,253 on the bottom
   left (red: 24$\mu$m, green: 8$\mu$m, blue: far-UV) and zoom into
   its central region (red: H$\alpha$, green: 20cm, blue:
   $0.2-1.5$\,keV, reproduced from \cite{Heesen2011}). Also shown is
   the CMZ of our Galaxy as seen by Bolocam, the VLA and Spitzer in
   the bottom right (Credit: A. Ginsburg, John Bally, F. Yusef-Zadeh,
   Bolocam GPS team; GLIMPSE II team). The CMZ of NGC\,253 and the
   Milky Way show striking similarities in $^{12}$CO$(2-1)$ line
   emission (see text for more details). Image reproduced from
   \cite{Sakamoto2011} .}
   \label{fig2}
\end{center}
\end{figure}

The synchrotron emission from starburst galaxies is in general
correlated to their far-infrared (FIR) luminosity and hence
star-formation rate (\cite{vanBuren1994}). The radio emission of
NGC\,253 is coincident with the $\sim$0.5~kpc central molecular zone
(CMZ), which has a star-formation rate exceeding that of the entire
MW. Fig.~\ref{fig2} provides a multi-frequency view of NGC~253 on the
scale of the galaxy as a whole and close to the nucleus. The nucleus
exhibits bright infra-red and radio emission, and an outflow is
apparent in thermal X-rays. In a recent CO study \cite{Sakamoto2011}
found striking morphological similarities between the CMZ of NGC\,253
and that of the MW (see Fig.~\ref{fig2}). The physical size of the CMZ
as a whole, and that of the prominent CO peaks ($20-50$\,pc), are
remarkably similar in the two systems. Although the molecular mass
estimate is afflicted by a considerable error in the CO-to-H$_2$
conversion ratio, the Giant Molecular Cloud (GMC) complexes in
NGC\,253 have significantly higher CO$(2-1)$ intensities, suggesting
order of magnitude higher densities in NGC\,253 and a factor 10 higher
total mass.

\begin{table}
  \begin{center}
    \caption{Comparison of NGC\,253 and both the CMZ of the Milky Way
      and our galaxy as a whole. GeV and TeV luminosities are quoted
      for 0.1-100 GeV and 0.1-100 TeV respectively.}
    \label{table1}
    \scriptsize{
      \begin{tabular}{l|c|c|c|c}\hline
        & CMZ NGC 253 & CMZ Milky Way & Milky Way & References \\\hline
        size (FWHM, kpc$^2$) & $\sim0.5 \times 0.1$ & $\sim0.5 \times 0.1$ & $\sim25\times 0.1$ & (1,2) \\ 
        $L_{\rm IR} (L_\odot)$ & $\sim3\times10^{10}$ & $\sim4\times10^{8}$ & $(1-3)\times 10^{10}$ & (1,3,4)\\
        $M_{\mathrm{gas}} (M_\odot)$ & $\sim5\times10^{8}$ & $\sim5\times10^{7}$ & $\sim(5-10)\times10^{9}$ & (1,5) \\
        SFR ($M_\odot$\,yr$^{-1}$) & 3.5$^*$ & 0.08 & $0.7-1.5$ &  (6,7,8)\\
       $\nu_{\mathrm{SN}}$ (yr$^{-1}$) & 0.03-0.08 & $\sim(2-8)\times 10^{-4}$ & $0.02\pm0.01$ & (9,10,11,12)\\
        \hline
        $L_{\rm GeV}(L_\odot)$ & $\sim3\times 10^6$ & $\sim400\ (2100)^\dagger$ & $\sim2\times 10^5$ & (13,14)\\
        $L_{\rm TeV}(L_\odot)$ & $\sim4\times 10^5$ & $\sim100\ (160)^\ddagger$ & $\sim8\times 10^3$ & (14,15)\\\hline
         $L_{\rm GeV}/\nu_{\mathrm{SN}}$ ($L_\odot$ Myr) & 40-100 &          0.5-10 & 6-20 & \\
         $L_{\rm TeV}/\nu_{\mathrm{SN}}$ ($L_\odot$ Myr) & 5-15 &         0.1-1 & 0.3-0.8 & \\ \hline
      \end{tabular}
    }
  \end{center}
  \vspace{1mm} \scriptsize{
    {\it References:}\\
    (1) \cite{Sakamoto2011}, (2) \cite{Combes1991}, (3) \cite{Sodroski1995}, 
    (4) \cite{Paladini2007}, (5) \cite{Pierce-Price2000}, (6) \cite{Melo2002}, 
    (7) \cite{Immer2011}, (8) \cite{Robitaille2010}, (9) \cite{vanBuren1994}, 
    (10) \cite{Engelbracht1998}, (11) \cite{Crocker2011}, (12) \cite{Diehl2006}, 
    (13) \cite{Fermi:NGC253M82}, (14) \cite{Strong2010} (15) \cite{HESS:GCR2006}\\
    {\it Notes:}\\
    $^*$ The star-formation rate has been inferred from the
    correlation with the $IR$ luminosity.\\
    $^\dag$ GeV luminosity of the CMZ as predicted by the \emph{Fermi}-LAT
    galactic diffuse model ({\it gal\_2yearp7v6\_v0.fits} from \url{http://fermi.gsfc.nasa.gov/ssc/data/access/lat/BackgroundModels.html}).
    Numbers in brackets include point-like sources from the Fermi 2-year catalogue (\cite{Fermi:2yr}) 
    in the region-of-interest. These numbers represent upper limits since they include 
    contributions of diffuse emission and unrelated point-like sources along the 
    line-of-sight.\\
    $^\ddag$ TeV luminosity of the CMZ as measured by H.E.S.S. (\cite{HESS:GCR2006}). 
    Numbers in brackets include the H.E.S.S. GC point-like source HESSJ1745--290 
    (\cite{HESS:GC2009}) and G\,0.9+0.1 (\cite{HESS:G09}).
  }
\end{table}

Table~\ref{table1} summarises SF and \g-ray emission related
quantities for the NGC\,253 starburst nucleus, for the MW CMZ and for
the MW as a whole. The SF rate of the MW and the CMZ are inferred from
number counts of young stellar objects and the application of a
stellar evolution model. The supernova rate of the CMZ has been
estimated by a number of different methods including the FIR-supernova
rate relation, stellar composition models, and pulsar population
studies (see \cite{Crocker2011} for a detailed discussion). Since the
distance to NGC\,253 is much larger than to the centre of our Galaxy,
young stellar objects are too faint to be detected by current
instruments, hence the SFR has to be inferred by indirect approaches
such as the scaling relation between SFR and IR flux
(\cite{Kennicutt1998}). The supernova rate also has to be indirectly
inferred, from e.g. iron-line measurements. The inferred SFR and
supernova rate in the starburst nucleus of NGC\,253 are orders of
magnitude larger than in the CMZ.

Under the assumption that particle acceleration predominantly occurs
in SNR shells, or is simply driven by massive star formation, and the
particle acceleration efficiency ($\epsilon_{\rm CR}$) is fixed, the
CR injection power should correlate with the supernova rate. For a
fixed efficiency of \g-ray production ($\epsilon_{\gamma}$) from
accelerated particles, the \g-ray flux should therefore be
proportional to the supernova rate.  Table~1 gives the ratio of the
\g-ray luminosity in the GeV and TeV bands to the supernova rate in
the three systems considered. Comparing the entire Milky Way with the
NGC\,253 starburst, a factor of a few difference apparently exists in
the product $\epsilon_{\rm CR}$$\epsilon_{\gamma}$.  A much more
dramatic difference exists between the apparently much more similar
regions of the NGC\,253 starburst and the CMZ, with $\epsilon_{\rm
  CR}$$\epsilon_{\gamma}$ smaller by 1-2 orders of magnitude with
respect to NGC\,253. This observation however, can be understood in
the following picture: The convective escape time, i.e. the residence
time of particles in both systems, is comparable if the CMZ in the MW
drives a nuclear wind, similar in terms of wind speed and size, to the
one observed in NGC\,253 (see e.g. \cite{Crocker2011}). The order of
magnitude higher density in the NGC\,253 starburst region on the other
hand, results in an order of magnitude higher fraction of energy lost
by an average proton in p-p interactions. Hence, the efficiency of
converting CR energy into \g\ rays ($\epsilon_{\gamma}$) is expected
to be an order of magnitude increased relative to the CMZ in the MW.

\section{Prospects for CTA}

The upcoming Cherenkov Telescope Array project (\cite{Actis2011})
represents a major step forward in both the sensitivity and precision
of \g-ray astronomy. Compared to current instruments it will provide a
broader energy coverage and overlap in energy with the \emph{Fermi}
satellite, a much better angular resolution, and a factor of 10 better
sensitivity. This will allow us to measure the spectra of M\,82 and
NGC\,253 in great detail and to search for spectral features such as
the emergence and subsequent disappearance of an IC component. If the
\g-ray emission from the known starbursts originates from the GMC
complex seen e.g. in CO, it might be possible to detect a significant
extension in the \g-ray emission, given the arc-minute-scale angular
resolution of CTA. Finally, the greatly improved sensitivity might
allow us to detect other nearby galaxies such as Andromeda (M\,31)
and/or to establish new source classes such as ultra-luminous infrared
galaxies, helping us to gain a deeper insight into the relationship
between star formation and particle acceleration.


\end{document}